\documentclass[aip,numerical]{revtex4-2}
\usepackage{graphicx}
\usepackage{siunitx}
\DeclareSIUnit{\Angstrom}{\textup{\AA}}
\usepackage{soul}

\usepackage[english]{babel}
\usepackage{blindtext}

\newcommand{\GeSi}{Ge$_{0.85}$Si$_{0.15}$ }
\newcommand{\gammaPoint}{$\Gamma$-point }
\newcommand{\LPoint}{L-point }
\usepackage{xcolor}
\usepackage{float}

\begin{document}

\title{Ge$_{1-x}$Si$_x$ single crystals for Ge hole spin qubit integration}

\author{Andreas Fuhrberg}
    \affiliation{Fachbereich Physik, Universität Konstanz, 78457 Konstanz, Germany}
\author{Pia M. Düring}
    \affiliation{Fachbereich Physik, Universität Konstanz, 78457 Konstanz, Germany}
\author{Olena Fedchenko}
	\affiliation{Institut für Physik, Johannes Gutenberg-Universität, 55128 Mainz, Germany}
    \affiliation{Physikalisches Institut, Goethe-Universität Frankfurt, 60438 Frankfurt am Main, Germany}
\author{Olena Tkach}
	\affiliation{Institut für Physik, Johannes Gutenberg-Universität, 55128 Mainz, Germany}
\author{Yaryna Lytvynenko}
    \affiliation{Institut für Physik, Johannes Gutenberg-Universität, 55128 Mainz, Germany}
    \affiliation{V.G. Baryakhtar Institute of Magnetism of the NAS of Ukraine, 03142 Kyiv, Ukraine}
 \author{Kevin Gradwohl}
    \affiliation{Leibniz Institut für Kristallzüchtung, 12489 Berlin, Germany}
\author{Sergii Chernov}
    \affiliation{Deutsches Elektronen-Synchrotron, 22607 Hamburg, Germany}
\author{Andrei Gloskovskii}
    \affiliation{Deutsches Elektronen-Synchrotron, 22607 Hamburg, Germany}
\author{Christoph Schlueter}
    \affiliation{Deutsches Elektronen-Synchrotron, 22607 Hamburg, Germany}
\author{Gerd Schönhense}
	\affiliation{Institut für Physik, Johannes Gutenberg-Universität, 55128 Mainz, Germany}
\author{Hans-Joachim Elmers}
    \affiliation{Institut für Physik, Johannes Gutenberg-Universität, 55128 Mainz, Germany}
\author{Martina Müller*}
    \affiliation{Fachbereich Physik, Universität Konstanz, 78457 Konstanz, Germany}
    \email{martina.mueller@uni-konstanz.de}
    
\date{\today{}}


\begin{abstract}
\section*{Abstract}
Spin qubits are fundamental building blocks of modern quantum computing devices. Among the various design approaches for spin qubits, the path of Ge-based hole-spin qubits has several advantages over Si-based electron-spin systems, such as the absence of valley band degeneracy, the possibility of efficient field control due to large spin-orbit coupling, and smaller effective masses.  Among the possible Ge qubit devices, Ge/GeSi planar heterostructures have proven to be favourable for upscaling and fabrication. The Si concentration of the straining GeSi buffer serves as an important tuning parameter for the electronic structure of Ge/GeSi qubits. A particularly low Si concentration of $x=0.15$ of the Ge$_{0.85}$Si$_{0.15}$ crystal should enable minimal lattice strain for spin qubit heterostructures, which is, however, difficult to stabilize as a random alloy.\\ 
We present a synchrotron-based study to investigate the chemical composition, valence band electronic structure and local atomic structure  of a Ge$_{0.85}$Si$_{0.15}$ single crystal using the advanced combination of hard X-ray photoelectron spectroscopy (HAXPES), hard X-ray momentum microscopy (HarMoMic) and X-ray photoelectron diffraction (XPD). We found that - compared to semiconductor grade pure Si and Ge wafers - the Ge$_{0.85}$Si$_{0.15}$ crystal has an individual, uniform valence band structure, with no signs of phase separation. The shapes of the heavy/light hole band and the split-off band resemble those of pure Ge, as do the low effective masses. XPD experiments, supported by Bloch wave calculations, show that the Si atoms are located at Ge lattice sites within the Ge$_{0.85}$Si$_{0.15}$ crystal, forming a random alloy.
This high chemical, electronic and structural quality of Ge$_{0.85}$Si$_{0.15}$ single-crystal substrates is of crucial importance for their implementation in hybrid quantum systems, in particular to enable long spin lifetimes in Ge-based hole-spin qubits. The results emphasise the power of combined X-ray spectromicroscopy techniques, which provide key insights into the qubit building blocks that form the basis of quantum technologies.
\end{abstract}

\maketitle 


\section{Introduction}
Semiconductor spin qubits have made significant advances in recent years, with the demonstration of universal control in quantum processors featuring up to six qubits \cite{Paper:Hendrickx2021,Paper:Philips2022}, and even 12-qubit arrays fabricated on a $\SI{300}{\milli\m}$ semiconductor manufacturing line \cite{Paper:George2024}. Current research is concentrated on improving the scalability of this technology.

In particular, germanium-based hole-spin qubits in strained Ge quantum wells have emerged as a promising platform for upscaling and fabrication. Quantum confined holes in Ge enable long spin lifetimes and long coherence times as well as immunity to the valley band degeneracy issues that affect electron spin qubits based on Si quantum wells \cite{Paper:Scappucci2020}. A straining Ge-rich GeSi buffer layer creates a valence-band offset at the interface and serves therefore as an important tuning parameter for the electronic structure of Ge/GeSi qubits. However, current qubit demonstrations are based on GeSi strain-relaxed buffer layers grown on Si wafers, which inherently suffer from a high concentration of crystal defects, compromising scalability \cite{Paper:Stehouwer2023}. 

Bulk GeSi single crystals present an alternative platform that promises significantly improved crystalline quality, particularly in terms of chemical homogeneity and dislocation density. These improvements could enable longer spin life- and coherence times, which are essential for scalable spin qubits \cite{Paper:Subramanian2025}. However, stabilizing high-quality bulk SiGe crystals with uniform composition remains challenging due to solubility differences and melt inhomogeneities. 

A single-crystal GeSi suitable for implementation in quantum technology should meet the high quality standards of semiconductor-grade Si and Ge. On the microscopic scale, a Ge$_{0.85}$Si$_{0.15}$ crystal should exhibit an individual valence band structure, indicating a macroscopically homogeneous and pure phase alloy.  At the atomic level, this in turn means that Si atoms are located at Ge lattice sites within the Ge$_{0.85}$Si$_{0.15}$ crystal.

Here, we present an advanced synchrotron-based study that investigates the chemical, electronic and structural properties of a Ge$_{0.85}$Si$_{0.15}$ single crystal. Hard X-ray photoelectron emission is analysed in spectrocopic (HAXPES), microscopic (HarMoMic) and diffraction (XPD) modes to collect relevant information about the chemical bonding and composition, the momentum-resolved distribution of electronic states at and near the Fermi level, and the local atomic environment of adjacent lattice sites. This complementing set of microscopic information on a Ge$_{0.85}$Si$_{0.15}$ single crystal confirms the formation of an uniform valence band structure with no signs of phase separation. As expected for a Ge-rich Ge$_{0.85}$Si$_{0.15}$ alloy, the heavy/light hole band and the split-off band are similar to those of pure Ge. The effective masses determined from the valence band (VB) curvatures are of the same order of magnitude as for pure Ge.
XPD experiments are consistent with Bloch wave calculations, which indicate that the Si atoms are positioned at Ge lattice sites. 

These results not only demonstrate the high chemical, electronic and structural quality of Ge$_{0.85}$Si$_{0.15}$ single-crystal substrates, which is essential for the integration of Ge-based hole-spin qubits, but also highlights the strength of a multi-technique approach based on X-ray photoelectron emission spectromicroscopies to enable future quantum technologies.


\subsection{Experimental Methods}
A Ge$_{1-x}$Si$_x$ bulk single crystal was grown by the Czochralski method from a pure Ge melt and with continuous Si supply. Details on the growth technique and further material characterisation are provided in the SI. The investigated sample is a (001) wafer piece with a miscut below 0.5$^{\circ}$ and an RMS roughness of $<0.5$ nm. The average Si concentration was determined to be $x = 0.16$ and the dislocation density as $\SI{1E-5}{\per\centi\m\squared}$. More information about characterization of the elemental concentration and crystal growth can be found in Subramanian et al. \cite{Paper:Subramanian2025}. Additionally, commercial Ge(001) and Si(001) single-crystalline wafers were used as references.

The Ge$_{0.85}$Si$_{0.15}$ single crystal was analysed using complementary hard X-ray photoelectron spectromicroscopy methods to determine the chemical properties, electronic structure and local atomic environment. The experiments were performed at the P22 beamline of PETRA III at DESY, Hamburg \cite{Paper:P22Beamline}. 

The chemical composition of the sample was investigated by hard X-ray photoelectron spectroscopy (HAXPES) \cite{Paper:Mueller2022}. Photoelectron emission spectra of the Ge 2s, Ge 2p, Ge 3s, Ge 3p, Si 2s and Si 2p core levels were recorded at two photon energies, $\SI{6}{\kilo\electronvolt}$ and $\SI{2.8}{\kilo\electronvolt}$, resulting in information depths (ID) of $\SI{24}{\nm}$ and $\SI{12}{\nm}$ \cite{Paper:SESSA} for bulk and more surface sensitive measurements, respectively.
Peaks were fitted using a model that consists of the sum of a Gaussian peak for the oxide component and a Voigt peak for the unoxidised components of the spectra.

The valence band structure of the samples was investigated using a time-of-flight momentum microscope (MM) \cite{Chernov2015,Paper:P22MM,Paper:Rosenberger2025}. 
Momentum-energy maps $I(k_x , k_y, E_B $) were recorded on samples cooled down to $\SI{30}{\K}$ by liquid He with an energy resolution of $100$\,meV using a post-monochromator \cite{Paper:P22MM}. The photon energy required to map a specific $k_z$ plane of the samples k-space was calculated via the concept of direct transitions during the photoemission process applied for k-space \cite{medjanik_direct_2017}. 

Using the same time-of-flight momentum microscope, the crystal structure analysis of the samples was performed by hard X-ray photoelectron diffraction (hXPD). Diffraction pattern of the Si 2p and Ge 3p core levels were recorded using a kinetic energy of $\SI{3203}{\electronvolt}$, ($\SI{3180}{\electronvolt}$) $[\SI{3167}{\electronvolt}]$. Experimental XPD maps are compared to state-of-the-art dynamical calculations using the Bloch-wave approach\cite{Winkelmann2008,Winkelmann2004}.

The recorded data in form of 3D-data arrays was analysed by the open-source programme ImageJ \cite{Paper:Schindelin2012}. The analysis steps include a fourfold symmetrisation around the $\Gamma$-point justified by the present symmetry of the sample crystals to enhance data density.
From the 3D momentum-energy maps $I(k_x , k_y, E_B) $, a $(k_X, E_B)$ section can be isolated. By choosing the direction from the $\Gamma$-point to the X-point, as shown in Fig.\,\ref{fig:VB_allSiGe}\,(a) by the $\Delta$ vector, the valence band dispersion along this vector was extracted.


\subsection{Results}


\subsubsection{Chemical composition of the GeSi single crystal}
\begin{figure}[H]
    \centering
    \includegraphics[width=0.8\textwidth]{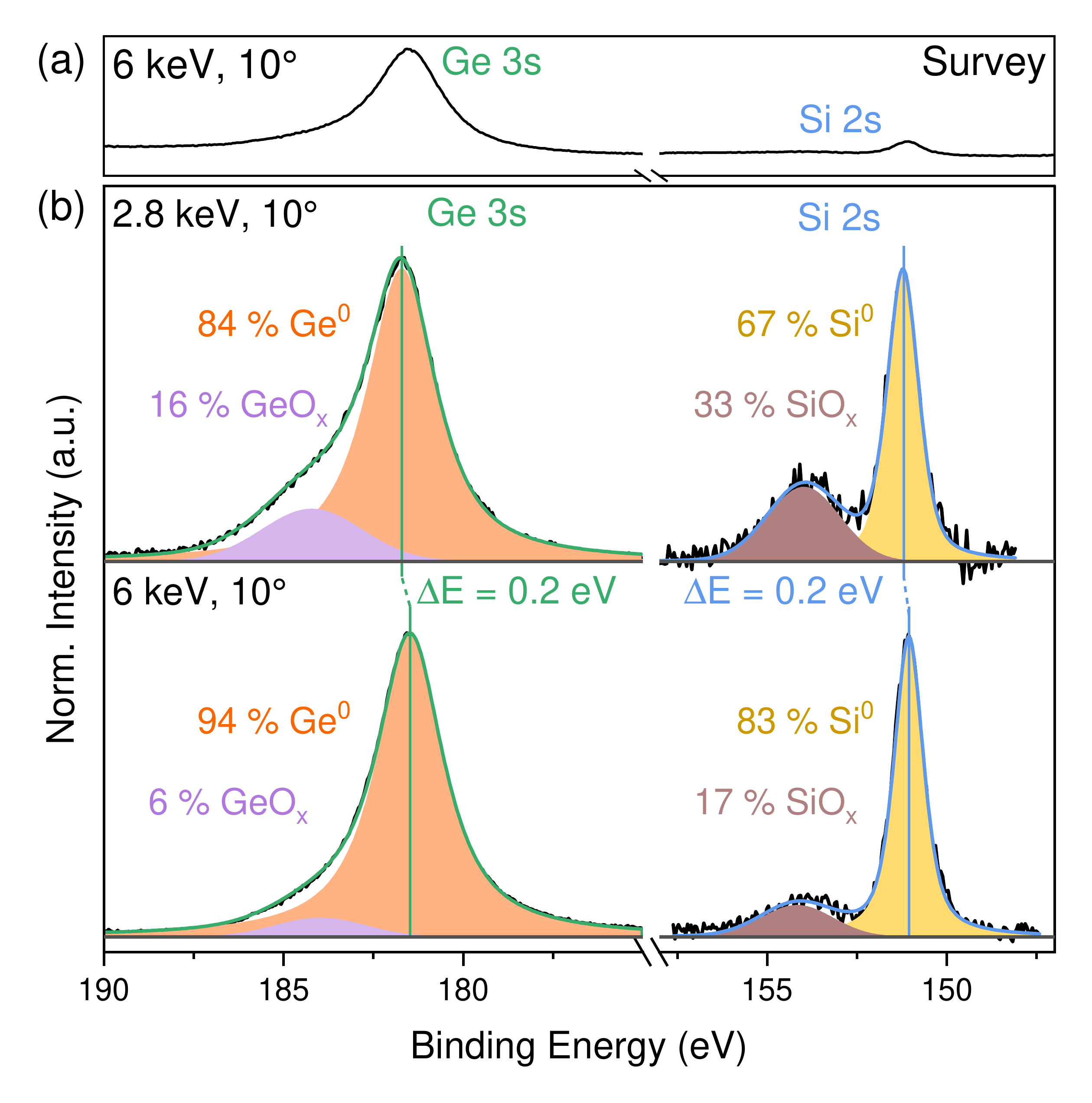}
    \caption{{\bf{Chemical properties of the GeSi single crystal determined by HAXPES.}} Photon energies of $\SI{2.8}{\kilo\electronvolt}$ and \SI{6}{\kilo\electronvolt} enable increased surface and bulk sensitivity, respectively. The Ge 3s (left) and Si 2s (right) core levels are shown, both of which exhibit slightly increased surface oxidation. Ge:Si composition analysis yields Ge$_{0.85}$Si$_{0.15}$ for $\SI{6}{\kilo\electronvolt}$.}
    \label{fig:HAXPES}
\end{figure}

An important parameter for tuning the electronic structure of strained Ge quantum wells is the Ge vs. Si concentration of the underlying GeSi template. The chemical composition of the Ge$_{0.85}$Si$_{0.15}$ single crystal was investigated by HAXPES. Figure\,\ref{fig:HAXPES} shows the Ge 2s and Si 2s core level spectra recorded at photon energies of $\SI{2.8}{\kilo\electronvolt}$ and $\SI{6}{\kilo\electronvolt}$. The survey scan in Fig.\,\ref{fig:HAXPES}\,(a) depicts the as-measured intensity ratio between the Ge and Si peaks, while the normalised core-level peaks are shown in Fig.\,\ref{fig:HAXPES}\,(b). Both an unoxidised and an oxide component appear in the Ge 3s and Si 2s spectra, respectively, located at lower and higher binding energies. Additional Information about the oxide components are given in the supplementary information (SI).

Based on an analysis of every recorded core level spectra of the elements Si and Ge, the chemical composition of the GeSi single crystal was determined. The Ge$_{1-x}$Si$_x$ stoichiometry is calculated from the cross section weighted peak areas \cite{Paper:Trzhaskovskaya}. This results in Ge$_{0.85}$Si$_{0.15}$ for $\SI{6}{\kilo\electronvolt}$ and Ge$_{0.8}$Si$_{0.2}$ for $\SI{2.8}{\kilo\electronvolt}$ within an accuracy of about $\SI{2}{\percent}$. Hence, the composition of Ge vs. Si varies only slightly over a range of $\SI{4}{\nm}$ to $\SI{5}{\nm}$ from the bulk towards the surface and does not change the overall electronic structure, which is investigated in the following.

\subsubsection{Electronic structure of Ge, Si and Ge$_{0.85}$Si$_{0.15}$ valence bands}
\begin{figure*}
    \centering
    \includegraphics[width=0.8\textwidth]{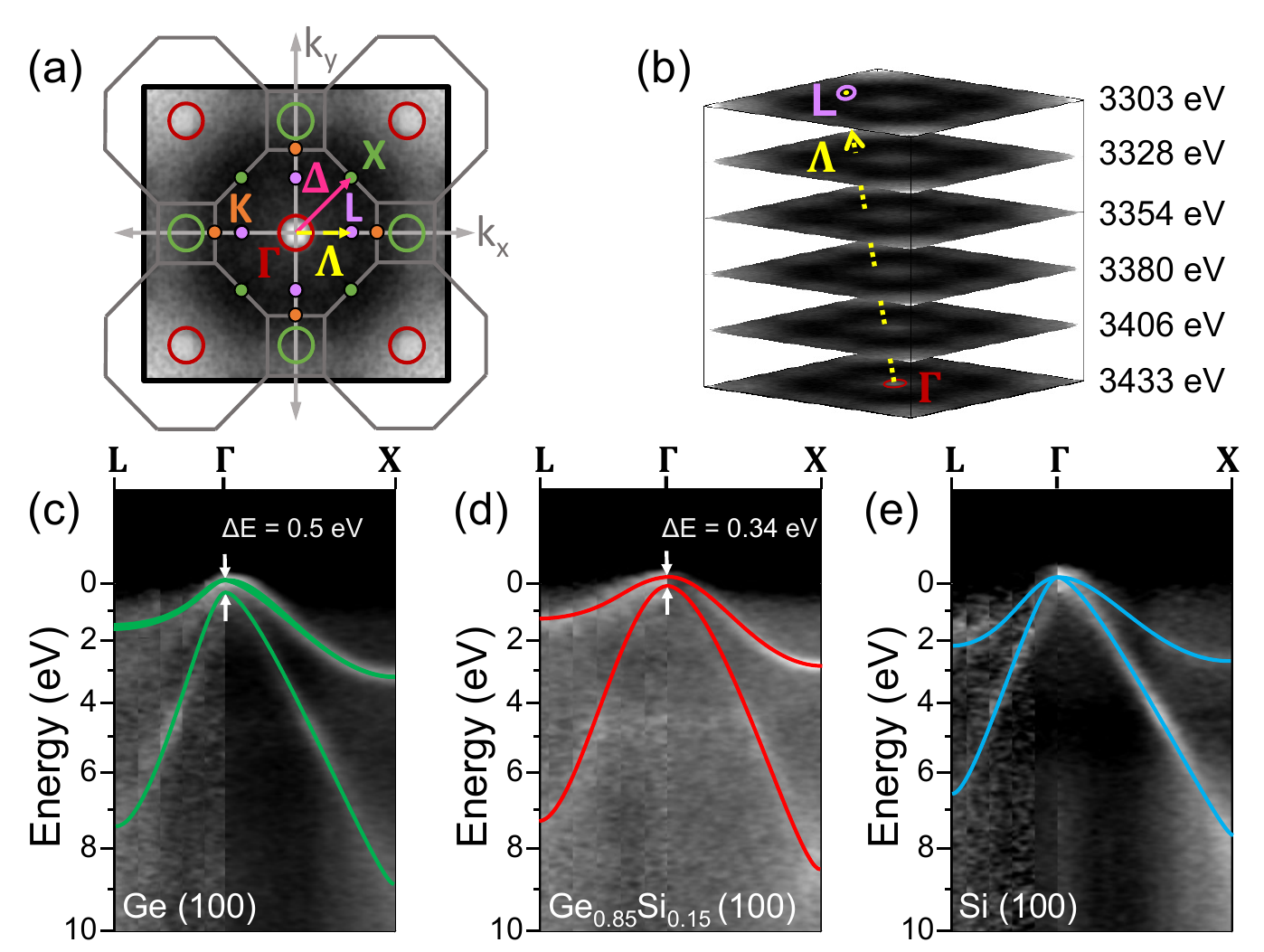}
    \caption{{\bf{Momentum-resolved photoelectron microscopy of Ge, Ge$_{0.85}$Si$_{0.15}$ and Si single crystalline samples.}} (a) $\Gamma$-plane of the Ge$_{0.85}$Si$_{0.15}$ sample. (b) Sliced $E(k)$ maps from the $\Gamma$- to the L-plane for Ge$_{0.85}$Si$_{0.15}$. (c)--(e) Valence band electronic structure of the Ge, Ge$_{0.85}$Si$_{0.15}$ and Si single crystal along $L-\Gamma - X$ direction.}
    \label{fig:VB_allSiGe}
\end{figure*}

In order to investigate the details of the valence band electronic structure of the Ge$_{0.85}$Si$_{0.15}$ single crystal, we performed momentum-resolved photoelectron microscopy experiments on (100) oriented single crystals. The photon energy was set to $\SI{3433}{\electronvolt}$, ($\SI{3140}{\electronvolt}$), and $[\SI{3420}{\electronvolt}]$ to select the $\Gamma$-planes of the Ge$_{0.85}$Si$_{0.15}$, (Ge), and $[\text{Si}]$ samples, respectively. The $\Gamma$-plane of the Ge$_{0.85}$Si$_{0.15}$ sample with the overlaid Brillouin zones is shown in Fig.\,\ref{fig:VB_allSiGe}\,(a).

In addition, the Brillouin zones (BZ) of the Si, Ge and Ge$_{0.85}$Si$_{0.15}$ samples were scanned by changing the photon energy. Thus, the perpendicular momentum $k_z$ was scanned between the $\Gamma$ and $L$ points in steps corresponding to about $0.1G_{001}$ [see e.g. Fig.\,\ref{fig:VB_allSiGe}\,(b)]. 
These photon energy scans result in photoemission intensities
$I(E_B,k_x,k_y)$ for $6$, $(8)$ and $[8]$ $k_z$ sections from the $\Gamma$-plane to the L-plane for the Ge$_{0.85}$Si$_{0.15}$, (Ge) and $[\text{Si}]$ sample, respectively. 
For Ge$_{0.85}$Si$_{0.15}$, the $E(k)$ sections are shown in Fig.\,\ref{fig:VB_allSiGe}\,(b). 
From the corresponding $I(E_B,k_x,k_y,k_z)$ data arrays, the 
band dispersion along the $\Gamma$ - L direction is interpolated.

In Fig.\,\ref{fig:VB_allSiGe}\,(c) and (e), the valence band structure of semiconductor-grade Ge and Si single crystals are shown from the L-point through the \gammaPoint towards the X-point. Guides to the eye are highlighted in green and blue, respectively. 

For the Si VB structure in Fig.\,\ref{fig:VB_allSiGe}\,(e), the VB maximum is located at the $\Gamma$-point and henceforth will serve as a reference energy for comparison with the Ge and Ge$_{0.85}$Si$_{0.15}$ samples. The light hole (LH) and heavy hole (HH) bands as well as the split-off (SO) band can be identified in the direction to the X- and L-points. Both bands show the characteristic non-parabolic shape. However, the substructure of the LH and HH bands, as well as the gap to the spin-orbit split-off (SO) band at the \gammaPoint can not be distinguished in our measurements: In Si, the SO gap has a magnitude of $\SI{0.044}{\electronvolt}$ \cite{Paper:Schäffler}, which is smaller than the energy resolution of the beamline \cite{Paper:P22MM} at the used photon energy. At the X-point, the LH/HH band has an energy minimum of $\SI{2.7 +- 0.1}{\electronvolt}$  and the SO band an energy minimum of $\SI{7.5 +- 0.1}{\electronvolt}$. These values agree with previously published data \cite{Paper:SiElectronicStructure}. 
Along the $\Gamma -$L direction the LH/HH band crosses the \LPoint at an energy of $\SI{2.2 +- 0.1}{\electronvolt}$ and the SO band at $\SI{6.9 +- 0.1}{\electronvolt}$. Using the curvature of the valence bands with a parabolic approximation  at the \gammaPoint , the effective hole masses $m^* _{\text{Si},LH/HH} =\SI{0.13 +- 0.1 }{} m_0 $ and $m^* _{\text{Si},SO} =\SI{0.08 +- 0.01 }{} m_0 $ are calculated as multiples of the free electron mass $m_0$. Due to resolution limits, these values represent a rough estimate, but the order of magnitude agrees with previously published values~\cite{Paper:SiElectronicStructure}.

Next, we turn to the results of the band structure measurement of the Ge crystal, shown in Fig.\,\ref{fig:VB_allSiGe}\,(a). In comparison to the Si VB, the splitting of the energy gap between SO and LH/HH band can be resolved. The SO split-off energy is determined as $\SI{0.5 +- 0.1}{\electronvolt}$. 

The LH/HH band of Ge has a broader shape compared to Si, which is due to the more pronounced variations of the LH and HH bands for Ge, as it is known from the literature \cite{Paper:Schäffler}. At the X-point the LH/HH band has an energy of $\SI{3.1 +- 0.1}{\electronvolt}$ and the SO band of $\SI{8.9 +- 0.1}{\electronvolt}$. The band energy at the \LPoint is $\SI{1.4 +- 0.1}{\electronvolt}$ for the LH/HH band and $\SI{7.5 +- 0.1}{\electronvolt}$ for the SO band. Similar to Si, the effective hole masses for both bands result in $m^* _{\text{Ge},LH/HH} =\SI{0.08 +- 0.01 }{}\, m_0 $ and $m^* _{\text{Ge},SO} =\SI{0.06 +- 0.02 }{}\, m_0 $.

Finally, the valence band structure of the Ge$_{0.85}$Si$_{0.15}$ crystal is shown in Fig.\,\ref{fig:VB_allSiGe}\,(d). Red lines serve as guides to the eye. The general shape of the bands are similar to the Ge reference crystal, as known in the literature for a Si concentration of $\SI{0.15}{}$ \cite{Paper:Schäffler}. The Ge$_{0.85}$Si$_{0.15}$ crystal has a smaller SO gap of $\SI{0.35 +- 0.1}{\electronvolt}$ as compared to Ge. Additionally, both the LH/HH and SO band have reduced binding energies at both symmetry points X and L. For the X-point, the LH/HH band has an energy of $\SI{2.96 +- 0.1}{\electronvolt}$ and the SO band of $\SI{8.75 +- 0.1}{\electronvolt}$. At the L-point, the bands have an energy of $\SI{1.37  +- 0.1}{\electronvolt}$ for the LH/HH band and $\SI{7.4 +- 0.1}{\electronvolt}$ for the SO band. The reduced binding energies at the X- and L-points result in a change in the slope and therefore in a change of the band curvature. A different curvature directly changes the effective masses of the band carriers. As a result, the effective hole masses of the  Ge$_{0.85}$Si$_{0.15}$ crystal change in comparison to the Ge crystal. A quantitative evaluation results in $m^* _{\text{GeSi},LH/HH} =\SI{0.07 +- 0.02 }{} m_0 $ and $m^* _{\text{GeSi},SO} =\SI{0.05 +- 0.03 }{} m_0 $. 

The main result of the comparison between the valence band electronic structure of semiconductor-grade Ge and Si samples and the Ge$_{0.85}$Si$_{0.15}$ single crystal is that the valence band mapping of the Ge$_{0.85}$Si$_{0.15}$ single crystal reveals a fully individual band structure. The overall shape still resembles that of a pure Ge-based valence band structure, which is however modified by alloying the crystal with Si. In addition, the effective masses determined from the momentum microscopy data are in the same order of magnitude and numerically close to the pure Ge and Si references. We therefore conclude that (i) the Ge$_{0.85}$Si$_{0.15}$ single crystal is electronically uniform, (ii) that the distribution of the alloyed Si in the Ge matrix is homogeneous, and (iii) that a phase segregation into Si and Ge rich areas can be excluded. Structural aspects will be elucidated in the following section.


\subsubsection*{Structural insights into Si, Ge and Ge$_{0.85}$Si$_{0.15}$ by X-ray photoelectron diffraction}

The goal of the XPD study is to investigate the structural properties of the GeSi system within the probed volume of the valence band momentum microscopy. XPD is performed at high resolution and small photoelectron collection angle. The measured diffraction patterns are then compared with Bloch wave calculations. In general, high-resolution $k$-space mapping (hXPD) provides increased sensitivity to the effects of long-range order, which manifest itself in the fine structure of bulk diffraction features. In contrast, wide-angular-range investigations focus on the local order around the photoemitter sites (even without long-range order), as manifested e.g. by the forward-focussing directions toward neighbour sites~\cite{Fedchenko2019}.
Thus, hXPD provides element and site-specific atomic order~\cite{Medjanik2021} and high-precision site locations~\cite{Hoesch2023}.

\begin{figure*}
    \centering
    \includegraphics[width=0.8\textwidth]{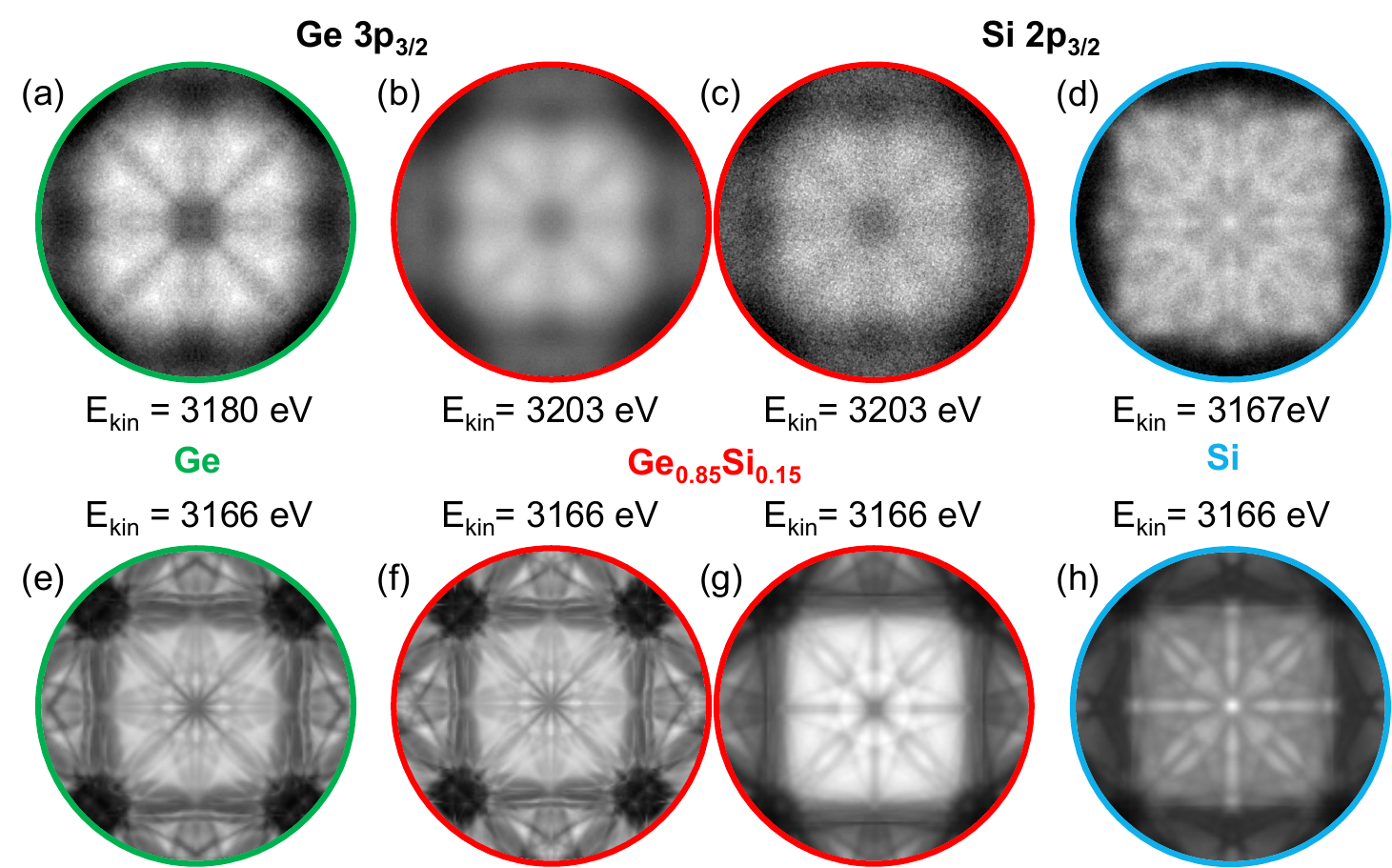}
    \caption{{\bf{High-resolution X-ray photoelecton diffraction pattern of Ge 3p$_{3/2}$ and Si 2p$_{3/2}$ core levels of the (a) Ge, (b),(c) Ge$_{0.85}$Si$_{0.15}$ and (d) Si single crystals.}} The photon energies were varied to achieve kinetic energies around $E_\text{kin}\approx\SI{3200}{\eV}$. (e)-(h) Simulated XPD core level spectra at $E_\text{kin}=\SI{3166}{\eV}$ using Bloch wave calculations.}\label{Figure3}
\end{figure*}

Figure\,\ref{Figure3} shows experimental (upper row) and calculated hXPD patterns (lower row, blurred by a Gaussian function to account for the experimental resolution) for the pure semiconductor single crystals Si and Ge and for the SiGe alloy. Photoelectrons were excited from the Si $2p$ and Ge $3p$ core levels. The kinetic energy of the electrons contributing to the XPD patterns was adjusted by the photon energy such that they are approximately the same. Then the wavelength of the emitted photoelectrons is similar and one would expect the same XPD pattern for Si and Ge if the scattering atoms could be approximated as point scatterers on the same lattice structure.

The experimental result show clear differences for the pure Si and Ge single crystals. In the case of Si, see Fig.\,\ref{Figure3}\,(d), a high intensity is observed for normal emission (central point) with linear bright stripes in the perpendicular and horizontal directions. Double stripes of high intensity occur in the diagonal directions. These features are in good agreement with the calculated results, as shown in Fig.\,\ref{Figure3}\,(h).

In contrast, the XPD patterns for Ge [Fig.\,\ref{Figure3}\,(a,e)] show dark lines of low intensity in the horizontal, vertical and diagonal directions. At normal emission, low intensity is observed. Experimental and calculated results also show a good agreement in this case.

Overall, we attribute the differences between the diffraction patterns of the Si and Ge hXPD patterns to the different lattice constants of the Si and Ge lattices ($\SI{5.43}{\Angstrom}$ and Ge $\SI{5.66}{\Angstrom}$) and to the different scattering phases of Si and Ge.

Finally, when examining the Ge$_{0.85}$Si$_{0.15}$ crystal, the XPD patterns recorded from the Ge $3p$ and Si $2p$ core levels have similar patterns as shown Fig.\,\ref{Figure3}\,(b,c,f,g). Both XPD patterns resemble those observed for the Ge reference and significantly deviate from the results obtained for the pure Si crystal. Therefore, we conclude that the Si and Ge atoms occupy the same lattice sites in the Ge lattice matrix of the Ge$_{0.85}$Si$_{0.15}$ crystal. In both cases, the Ge reference and the Ge$_{0.85}$Si$_{0.15}$ crystal, the scattering planes are mainly composed by Ge atoms.


\section{Conclusion}
In summary, we have investigated a Ge$_{0.85}$Si$_{0.15}$ single crystal with a particularly low Si concentration, which allows minimal lattice strain when implemented in spin qubit heterostructures.
In a comprehensive approach based on hard X-ray synchrotron radiation, we have evaluated the chemical composition, valence band electronic structure and structural diffraction pattern of Ge$_{0.85}$Si$_{0.15}$ using photoelectron spectroscopy, -microscopy and -diffraction.

The direct comparison between semiconductor-grade Ge and Si references and the Ge$_{0.85}$Si$_{0.15}$ single crystal shows that the valence bands of the Ge$_{0.85}$Si$_{0.15}$ single crystal exhibit an individual band structure. The overall shape is similar to that of a pure Ge-based valence band structure, but has been modified by alloying the crystal alloyed with $15\%$ Si. 

In particular, the effective masses determined from the momentum microscopy data are of the same order of magnitude and numerically close to the pure Ge and Si references. We conclude that the Ge$_{0.85}$Si$_{0.15}$ single crystal is electronically uniform, that the distribution of the alloyed Si in the Ge matrix is homogeneous, and that a phase segregation into Si and Ge rich areas can be ruled out. This findings are supported by the XPD results. The good agreement of experimental and calculated diffraction pattern also excludes the presence of a short range order of Si atoms in the Ge matrix. Consequently, the GeSi system forms a random alloy with a unique electronic band structure. 

This advanced approach using X-ray spectromicroscopy provides fundamental microscopic insights into the quantum material Ge$_{0.85}$Si$_{0.15}$. Furthermore, it demonstrates that such high chemical, electronic and structural quality of Ge$_{0.85}$Si$_{0.15}$ single crystalline substrates is essential for implementation in Ge-based quantum technology devices with long coherence and hole spin lifetimes. Upcoming electronic structure studies will include planar thin-film Ge/Ge$_{0.85}$Si$_{0.15}$ heterostructures, which serve as real-word building blocks for Ge-based spin qubits.

\begin{acknowledgments}
    We acknowledge insightful discussions with Guido Burkard at University of Konstanz. This work was supported by the Deutsche Forschungsgemeinschaft through Sonderforschungsbereich SFB 1432 (Project No. 425217212, Subproject No. B03), the Transregio TRR 173 (Project No. 268565370, Subproject No. A02), the Vector Foundation (project ID "iOSMEMO"), and by the German Federal Ministry of Education and Research under framework program ErUM (project 05K22VL1 and 05K22UM1). We acknowledge DESY (Hamburg, Germany), a member of the Helmholtz Association HGF, for the provision of experimental facilities. Parts of this research were carried out at PETRA III using beamline P22. Beamtime was allocated for proposals R-20240665, II-2023-0021 and I-20230526. 
\end{acknowledgments}

\bibliographystyle{naturemag}
\bibliography{GeSi_Bibliography.bib}

\begin{thebibliography}{10}
\expandafter\ifx\csname url\endcsname\relax
  \def\url#1{\texttt{#1}}\fi
\expandafter\ifx\csname urlprefix\endcsname\relax\def\urlprefix{URL }\fi
\providecommand{\bibinfo}[2]{#2}
\providecommand{\eprint}[2][]{\url{#2}}

\bibitem{Paper:Hendrickx2021}
\bibinfo{author}{Hendrickx, N.~W.} \emph{et~al.}
\newblock \bibinfo{title}{A four-qubit germanium quantum processor}.
\newblock \emph{\bibinfo{journal}{Nature}} \textbf{\bibinfo{volume}{591}},
  \bibinfo{pages}{580–585} (\bibinfo{year}{2021}).
\newblock \urlprefix\url{http://dx.doi.org/10.1038/s41586-021-03332-6}.

\bibitem{Paper:Philips2022}
\bibinfo{author}{Philips, S. G.~J.} \emph{et~al.}
\newblock \bibinfo{title}{Universal control of a six-qubit quantum processor in
  silicon}.
\newblock \emph{\bibinfo{journal}{Nature}} \textbf{\bibinfo{volume}{609}},
  \bibinfo{pages}{919–924} (\bibinfo{year}{2022}).
\newblock \urlprefix\url{http://dx.doi.org/10.1038/s41586-022-05117-x}.

\bibitem{Paper:George2024}
\bibinfo{author}{George, H.~C.} \emph{et~al.}
\newblock \bibinfo{title}{12-spin-qubit arrays fabricated on a 300 mm
  semiconductor manufacturing line}.
\newblock \emph{\bibinfo{journal}{Nano Letters}} \textbf{\bibinfo{volume}{25}},
  \bibinfo{pages}{793–799} (\bibinfo{year}{2024}).
\newblock \urlprefix\url{http://dx.doi.org/10.1021/acs.nanolett.4c05205}.

\bibitem{Paper:Scappucci2020}
\bibinfo{author}{Scappucci, G.} \emph{et~al.}
\newblock \bibinfo{title}{The germanium quantum information route}.
\newblock \emph{\bibinfo{journal}{Nature Reviews Materials}}
  \textbf{\bibinfo{volume}{6}}, \bibinfo{pages}{926--943}
  (\bibinfo{year}{2020}).
\newblock \urlprefix\url{https://www.nature.com/articles/s41578-020-00262-z}.

\bibitem{Paper:Stehouwer2023}
\bibinfo{author}{Stehouwer, L. E.~A.} \emph{et~al.}
\newblock \bibinfo{title}{Germanium wafers for strained quantum wells with low
  disorder}.
\newblock \emph{\bibinfo{journal}{Applied Physics Letters}}
  \textbf{\bibinfo{volume}{123}} (\bibinfo{year}{2023}).
\newblock \urlprefix\url{http://dx.doi.org/10.1063/5.0158262}.

\bibitem{Paper:Subramanian2025}
\bibinfo{author}{Subramanian, A.~N.} \emph{et~al.}
\newblock \bibinfo{title}{On the {Czochralski} growth of {Si$_{x}$Ge$_{1-x}$}
  crystals as substrates for strained {ge} quantum well heterostructures}.
\newblock \emph{\bibinfo{journal}{Journal of Applied Physics}}
  \textbf{\bibinfo{volume}{137}} (\bibinfo{year}{2025}).
\newblock \urlprefix\url{http://dx.doi.org/10.1063/5.0238533}.

\bibitem{Paper:P22Beamline}
\bibinfo{author}{Schlueter, C.} \emph{et~al.}
\newblock \bibinfo{title}{The new dedicated {HAXPES} beamline {P22} at
  {PETRAIII}}.
\newblock \emph{\bibinfo{journal}{AIP Conference Proceedings}}
  \textbf{\bibinfo{volume}{2054}}, \bibinfo{pages}{040010}
  (\bibinfo{year}{2019}).

\bibitem{Paper:Mueller2022}
\bibinfo{author}{Müller, M.} \emph{et~al.}
\newblock \bibinfo{title}{Hard x-ray photoelectron spectroscopy of tunable
  oxide interfaces}.
\newblock \emph{\bibinfo{journal}{Journal of Vacuum Science and Technology A}}
  \textbf{\bibinfo{volume}{40}}, \bibinfo{pages}{013215}
  (\bibinfo{year}{2022}).
\newblock \urlprefix\url{http://dx.doi.org/10.1116/6.0001491}.

\bibitem{Paper:SESSA}
\bibinfo{author}{Werner, W.}, \bibinfo{author}{Smekal, W.} \&
  \bibinfo{author}{Powell, C.~J.}
\newblock \emph{\bibinfo{title}{Simulation of Electron Spectra for Surface
  Analysis {(SESSA)} Version 2.2.2 User's Guide}} (\bibinfo{year}{2024}).
\newblock \urlprefix\url{https://doi.org/10.6028/NIST.NSRDS.100-2024}.

\bibitem{Chernov2015}
\bibinfo{author}{Chernov, S.} \emph{et~al.}
\newblock \bibinfo{title}{Anomalous d-like surface resonances on {Mo(110)}
  analyzed by time-of-flight momentum microscopy}.
\newblock \emph{\bibinfo{journal}{Ultramicroscopy}}
  \textbf{\bibinfo{volume}{159}}, \bibinfo{pages}{453--463}
  (\bibinfo{year}{2015}).

\bibitem{Paper:P22MM}
\bibinfo{author}{Medjanik, K.} \emph{et~al.}
\newblock \bibinfo{title}{Progress in {HAXPES} performance combining full-field
  {\it k}-imaging with time-of-flight recording}.
\newblock \emph{\bibinfo{journal}{Journal of Synchrotron Radiation}}
  \textbf{\bibinfo{volume}{26}}, \bibinfo{pages}{1996--2012}
  (\bibinfo{year}{2019}).
\newblock \urlprefix\url{https://doi.org/10.1107/S1600577519012773}.

\bibitem{Paper:Rosenberger2025}
\bibinfo{author}{Rosenberger, P.} \emph{et~al.}
\newblock \bibinfo{title}{Spin polarization of the two-dimensional electron gas
  at the {EuO/SrTiO$_3 $ } interface} (\bibinfo{year}{2024}).
\newblock \bibinfo{note}{Preprint at \url{https://arxiv.org/abs/2410.23804}}.

\bibitem{medjanik_direct_2017}
\bibinfo{author}{Medjanik, K.} \emph{et~al.}
\newblock \bibinfo{title}{Direct {3D} mapping of the {Fermi} surface and
  {Fermi} velocity}.
\newblock \emph{\bibinfo{journal}{Nature Materials}}
  \textbf{\bibinfo{volume}{16}}, \bibinfo{pages}{615--621}
  (\bibinfo{year}{2017}).
\newblock \urlprefix\url{https://www.nature.com/articles/nmat4875}.

\bibitem{Winkelmann2008}
\bibinfo{author}{Winkelmann, A.}, \bibinfo{author}{Fadley, C.~S.} \&
  \bibinfo{author}{Garcia~de Abajo, F.~J.}
\newblock \bibinfo{title}{High-energy photoelectron diffraction: model
  calculations and future possibilities}.
\newblock \emph{\bibinfo{journal}{New Journal of Physics}}
  \textbf{\bibinfo{volume}{10}}, \bibinfo{pages}{113002}
  (\bibinfo{year}{2008}).

\bibitem{Winkelmann2004}
\bibinfo{author}{Winkelmann, A.}, \bibinfo{author}{Schröter, B.} \&
  \bibinfo{author}{Richter, W.}
\newblock \bibinfo{title}{Simulation of high energy photoelectron diffraction
  using many-beam dynamical {Kikuchi-band} theory}.
\newblock \emph{\bibinfo{journal}{Physical Review B}}
  \textbf{\bibinfo{volume}{69}}, \bibinfo{pages}{245417}
  (\bibinfo{year}{2004}).

\bibitem{Paper:Schindelin2012}
\bibinfo{author}{Schindelin, J.} \emph{et~al.}
\newblock \bibinfo{title}{Fiji: an open-source platform for biological-image
  analysis}.
\newblock \emph{\bibinfo{journal}{Nature Methods}}
  \textbf{\bibinfo{volume}{9}}, \bibinfo{pages}{676–682}
  (\bibinfo{year}{2012}).
\newblock \urlprefix\url{http://dx.doi.org/10.1038/nmeth.2019}.

\bibitem{Paper:Trzhaskovskaya}
\bibinfo{author}{Trzhaskovskaya, M.} \& \bibinfo{author}{Yarzhemsky, V.}
\newblock \bibinfo{title}{Dirac–fock photoionization parameters for {HAXPES}
  applications}.
\newblock \emph{\bibinfo{journal}{Atomic Data and Nuclear Data Tables}}
  \textbf{\bibinfo{volume}{119}}, \bibinfo{pages}{99--174}
  (\bibinfo{year}{2018}).
\newblock
  \urlprefix\url{https://www.sciencedirect.com/science/article/pii/S0092640X16300596}.

\bibitem{Paper:Schäffler}
\bibinfo{author}{Schäffler, F.}
\newblock \bibinfo{title}{High-mobility si and ge structures}.
\newblock \emph{\bibinfo{journal}{Semiconductor Science and Technology}}
  \textbf{\bibinfo{volume}{12}}, \bibinfo{pages}{1515} (\bibinfo{year}{1997}).
\newblock \urlprefix\url{https://dx.doi.org/10.1088/0268-1242/12/12/001}.

\bibitem{Paper:SiElectronicStructure}
\bibinfo{author}{Chelikowsky, J.~R.} \& \bibinfo{author}{Cohen, M.~L.}
\newblock \bibinfo{title}{Electronic structure of silicon}.
\newblock \emph{\bibinfo{journal}{Phys. Rev. B}} \textbf{\bibinfo{volume}{10}},
  \bibinfo{pages}{5095--5107} (\bibinfo{year}{1974}).
\newblock \urlprefix\url{https://link.aps.org/doi/10.1103/PhysRevB.10.5095}.

\bibitem{Fedchenko2019}
\bibinfo{author}{Fedchenko, O.} \emph{et~al.}
\newblock \bibinfo{title}{High-resolution hard-x-ray photoelectron diffraction
  in a momentum microscope—the model case of graphite}.
\newblock \emph{\bibinfo{journal}{New Journal of Physics}}
  \textbf{\bibinfo{volume}{21}}, \bibinfo{pages}{113031}
  (\bibinfo{year}{2019}).

\bibitem{Medjanik2021}
\bibinfo{author}{Medjanik, K.} \emph{et~al.}
\newblock \bibinfo{title}{Site-specific atomic order and band structure
  tailoring in the diluted magnetic semiconductor {(In,Ga,Mn)As}}.
\newblock \emph{\bibinfo{journal}{Physical Review B}}
  \textbf{\bibinfo{volume}{103}}, \bibinfo{pages}{075107}
  (\bibinfo{year}{2021}).

\bibitem{Hoesch2023}
\bibinfo{author}{Hoesch, M.} \emph{et~al.}
\newblock \bibinfo{title}{Active sites of {Te-hyperdoped} silicon by hard x-ray
  photoelectron spectroscopy}.
\newblock \emph{\bibinfo{journal}{Applied Physics Letters}}
  \textbf{\bibinfo{volume}{122}} (\bibinfo{year}{2023}).

\bibitem{Paper:Luo2007}
\bibinfo{author}{Luo, Y.-R.}
\newblock \emph{\bibinfo{title}{Comprehensive Handbook of Chemical Bond
  Energies}} (\bibinfo{publisher}{CRC Press}, \bibinfo{year}{2007}).
\newblock \urlprefix\url{http://dx.doi.org/10.1201/9781420007282}.

\bibitem{Paper:Bera2005}
\bibinfo{author}{Bera, M.~K.} \emph{et~al.}
\newblock \bibinfo{title}{Rapid thermal oxidation of {Ge-rich}
  {Si$_{1-x}$Ge$_{x}$} heterolayers}.
\newblock \emph{\bibinfo{journal}{Journal of Vacuum Science {\&} Technology A:
  Vacuum, Surfaces, and Films}} \textbf{\bibinfo{volume}{24}},
  \bibinfo{pages}{84–90} (\bibinfo{year}{2005}).
\newblock \urlprefix\url{http://dx.doi.org/10.1116/1.2137329}.

\bibitem{Paper:Binder}
\bibinfo{author}{Binder, J.~F.}, \bibinfo{author}{Broqvist, P.},
  \bibinfo{author}{Komsa, H.-P.} \& \bibinfo{author}{Pasquarello, A.}
\newblock \bibinfo{title}{Germanium core-level shifts at {Ge/GeO$_{2}$}
  interfaces through hybrid functionals}.
\newblock \emph{\bibinfo{journal}{Phys. Rev. B}} \textbf{\bibinfo{volume}{85}},
  \bibinfo{pages}{245305} (\bibinfo{year}{2012}).
\newblock \urlprefix\url{https://link.aps.org/doi/10.1103/PhysRevB.85.245305}.

\end{thebibliography}

\newpage
\section{Supplementary information}
\subsection{Crystal growth and pre-characterisation}
A SiGe bulk crystal growth technique was developed at Leibniz Institut f\"ur Kristall\-z\"uch\-tung (Berlin, Germany) that replaces SiGe strain-relaxed buffer  layers with bulk wafers. Ge$_{1-x}$Si$_x$ bulk single crystals were grown using the Czochralski method with continuous Si feeding. A $[100]$ Ge seed was dipped into a pure Ge melt, followed by the gradual introduction of Si using symmetrically positioned Si rods. Counter-rotation of the crucible and crystal enhanced mixing and minimized the boundary layer. Growth was performed in an Ar atmosphere, with precise temperature control to regulate Si dissolution. The pulling and crucible translation rates were $\SI{1.5}{\milli\m\per\hour}$ and $\SI{0.5}{\milli\m\per\hour}$, respectively, yielding a net crystal growth rate of approximately $\SI{1}{\milli\m\per\hour}$. Prior to growth, Si rods and Ge feedstock underwent wet-chemical etching. 

The investigated $(0 0 1)$ SiGe wafer with an off-cut below 0.5° was chemo-mechanically polished to an RMS-roughness below $\SI{0.5}{\nano\m}$. It was extracted from the cylindrical section of the crystal with a constant composition, an average Si concentration of  $x = 0.16$ and a dislocation density of around $\SI{1E5}{\per\cm\squared}$  in the center.

A wafer was extracted from a bulk-grown Ge$_{1-x}$Si$_x$ crystal and subjected to further analysis of the currently best SiGe bulk wafer with a concentration of $x = 0.16$ and a dislocation density below $\SI{1E6}{\per\cm\squared}$ . The structural and electronic properties of the wafer were characterized to assess the material’s suitability for quantum applications. For more details see Ref.\cite{Paper:Subramanian2025}.

\subsection{Oxidation of the Si, Ge and \GeSi samples}

\begin{figure}[H]
    \centering
    \includegraphics[width=0.8\textwidth]{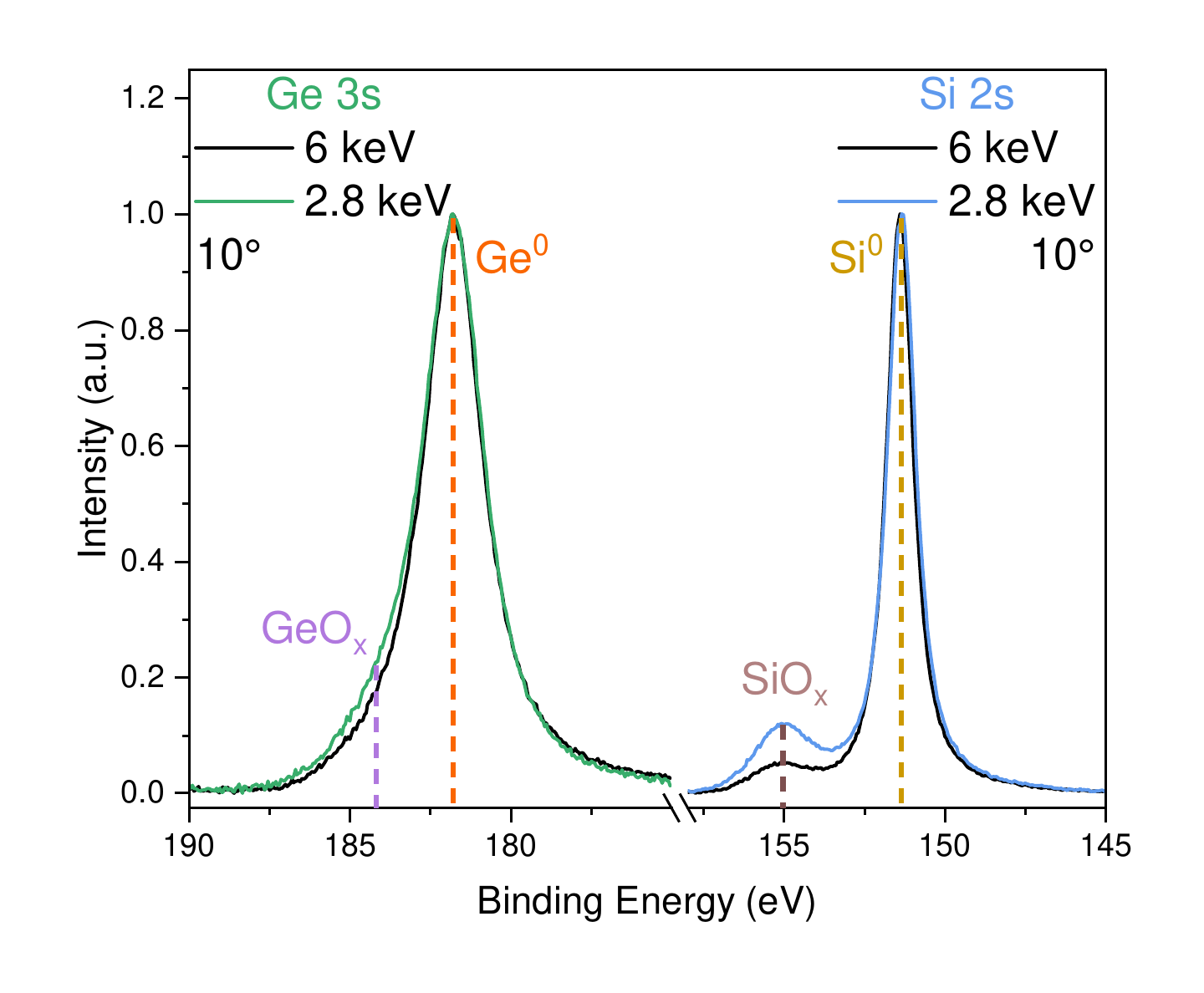}
    \caption{Oxidation state of the plain Ge and Si substrates, determined by HAXPES. Photon energies of $\SI{2.8}{\kilo\electronvolt}$ and $\SI{6}{\kilo\electronvolt}$ enable increased surface and bulk sensitivity, respectively. The Ge 3s (left) and Si 2s (right) core levels are shown, both of which exhibit slightly increased oxidation in the more surface near area of the samples. The magnitude of the oxidation is slightly increased for Si compared to Ge, due to the larger enthalpic energy gain during oxide formation.}
    \label{fig:HAXPES_Ge_Si_Oxidation}
\end{figure}

Besides the composition analysis, hard X-ray photoelectron spectroscopy (HAXPES) also provides information about the oxidation state of each component, as seen in Fig.\,\ref{fig:HAXPES} in the main article. In the more surface sensitive spectra recorded at $\SI{2.8}{\kilo\electronvolt}$, the GeO$_x$ and SiO$_x$ components are enhanced by $\SI{10}{\percent}$ and $\SI{16}{\percent}$ compared to the more bulk sensitive spectra at $\SI{6}{\kilo\electronvolt}$, respectively. In Figure \ref{fig:HAXPES_Ge_Si_Oxidation}, complementary HAXPES measurements on the plain Ge and Si substrates also show that the oxidation of the Si component is favoured compared to Ge. In fact, SiO$_x$ is preferentially formed compared to GeO$_x$, which is evident from the enthalpy of formation and bond strength under normal conditions and for Ge-rich GeSi alloys \cite{Paper:Luo2007, Paper:Bera2005}.

A rigid binding energy shift of $\SI{0.2}{\electronvolt}$ towards higher binding energy is observed between the surface and bulk sensitive measurements. This rigid shift can be explained as a result of the increased oxide contribution in the $\SI{2.8}{\kilo\electronvolt}$ measurements \cite{Paper:Binder}.

Since the oxidised parts of the samples are amorphous in nature, they do not contribute to the k-resolved valence band structure detected by momentum microscopy, but add to the background signal. 
Therefore, the oxidation has no impact on the valence band structure measurements via the momentum microscopy method.

\end{document}